\documentclass[
reprint,
floatfix,
superscriptaddress,
amsmath,
amssymb,
aps,
prl,
noeprint,
]{revtex4-2}

\usepackage{graphicx}
\usepackage{xcolor}
\usepackage{bm}
\usepackage{hyperref}
\hypersetup{colorlinks,linkcolor=[rgb]{0,0,0.5},citecolor=[rgb]{0,0,0.5},urlcolor=[rgb]{0,0,0.5}}
\usepackage{physics} 

\newcommand{\resultgraphic}[1]{%
  \IfFileExists{#1}{\includegraphics[width=\linewidth]{#1}}{%
    \fbox{\parbox[c][0.48\linewidth][c]{0.92\linewidth}{\centering\small
    Figure placeholder\\[0.3em]\texttt{\detokenize{#1}}}}}}
\newcommand{\resultgraphicat}[2]{%
  \IfFileExists{#2}{\includegraphics[width=#1]{#2}}{%
    \fbox{\parbox[c][0.30\textwidth][c]{0.48\textwidth}{\centering\small
    Figure placeholder\\[0.3em]\texttt{\detokenize{#2}}}}}}

\begin{document}

\title{Nucleon transversity from dihadron data and \texorpdfstring{$x$}{x}-dependent lattice QCD constraint}

\author{Hongxin Dong}
\affiliation{Nanjing Normal University, Nanjing, Jiangsu 210023, China}
\affiliation{Institute of Modern Physics, Chinese Academy of Sciences,
Lanzhou, Gansu 730000, China}

\author{Yuan-Yuan Huang}
\affiliation{Nanjing Normal University, Nanjing, Jiangsu 210023, China}

\author{Tianbo Liu}
\email[Corresponding author: ]{liutb@sdu.edu.cn}
\affiliation{Key Laboratory of Particle Physics and Particle Irradiation
(MOE), Institute of Frontier and Interdisciplinary Science, Shandong
University, Qingdao, Shandong 266237, China}
\affiliation{Southern Center for Nuclear-Science Theory (SCNT), Institute
of Modern Physics, Chinese Academy of Sciences, Huizhou, Guangdong
516000, China}

\author{Xiaomin Shen}
\affiliation{Institute of Modern Physics, Chinese Academy of Sciences,
Lanzhou, Gansu 730000, China}
\affiliation{University of Chinese Academy of Sciences, Beijing 100049,
China}

\author{Peng Sun}
\email[Corresponding author: ]{pengsun@impcas.ac.cn}
\affiliation{Institute of Modern Physics, Chinese Academy of Sciences,
Lanzhou, Gansu 730000, China}
\affiliation{University of Chinese Academy of Sciences, Beijing 100049,
China}

\author{Xiaonu Xiong}
\email[Corresponding author: ]{xnxiong@csu.edu.cn}
\affiliation{School of Physics and Electronics, Central South University,
Changsha 418003, China}

\author{Jian-Hui Zhang}
\email[Corresponding author: ]{zhangjianhui@cuhk.edu.cn}
\affiliation{School of Science and Engineering, The Chinese University of Hong Kong (Shenzhen), Longgang, Shenzhen, Guangdong, 518172, P.R. China}
\affiliation{Southern Center for Nuclear-Science Theory (SCNT), Institute of Modern Physics, Chinese Academy of Sciences, Huizhou, Guangdong 516000, China}

\author{Yuxiang Zhao}
\email[Corresponding author: ]{yxzhao@impcas.ac.cn}
\affiliation{Southern Center for Nuclear-Science Theory (SCNT), Institute
of Modern Physics, Chinese Academy of Sciences, Huizhou, Guangdong
516000, China}
\affiliation{Institute of Modern Physics, Chinese Academy of Sciences,
Lanzhou, Gansu 730000, China}
\affiliation{University of Chinese Academy of Sciences, Beijing 100049,
China}
\affiliation{Institute of Particle Physics, Central China Normal University,
Wuhan 480079, China}

\author{Yiyu Zhou}
\affiliation{Institute of Modern Physics, Chinese Academy of Sciences,
Lanzhou, Gansu 730000, China}
\affiliation{University of Chinese Academy of Sciences, Beijing 100049,
China}

\author{Ruilin Zhu}
\affiliation{Nanjing Normal University, Nanjing, Jiangsu 210023, China}

\collaboration{Transverse Nucleon Tomography Collaboration}

\begin{abstract}
We extract collinear transversity distributions and dihadron fragmentation functions in a joint neural-network analysis of dihadron production in electron–positron annihilation, semi-inclusive deep-inelastic scattering, and polarized proton–proton collisions. We incorporate lattice QCD calculation of isovector transversity, which constrains the up–down flavor differences for quarks and antiquarks over broad momentum fraction range complementary to data. Matched fits with and without the lattice input describe the experimental data comparably well. Including the lattice input shifts the down-valence distribution toward more negative values and enhances up-valence transversity at large $x$. The magnitude of the central down-valence result exceeds the reference Soffer inequalities, motivating further test of positivity constraints.
\end{abstract}

\maketitle

Understanding the quark and gluon structure of nucleons is a central goal of quantum chromodynamics (QCD). Parton distribution functions (PDFs) describe the internal structure in terms of the momentum fraction $x$ carried by partons in the nucleon in high-energy scattering~\cite{Aidala:2012mv,Gao:2017yyd}. When nucleon and parton spins are included, one can define three collinear PDFs at the leading twist: the unpolarized distribution $f_1$, the helicity distribution $g_1$, and the transversity distribution $h_1$. Transversity measures the difference between quark densities with transverse spin aligned and antialigned with that of a transversely polarized nucleon~\cite{Ralston:1979ys,Artru:1989zv,Jaffe:1991ra,Barone:2001sp}. Unlike the helicity distribution, transversity is a chiral-odd quantity, hence it is dominated by the valence nature and does not mix with gluons. However, this property also prevents the access through inclusive deep-inelastic scattering at leading twist, making it the least known among nucleon collinear PDFs. 

Experimental access to transversity distributions requires convoluting to another chiral-odd quantity. The Collins effect in single-hadron semi-inclusive deep-inelastic scattering (SIDIS) provides one route through the transverse-momentum-dependent (TMD) factorization~\cite{Collins:1992kk}. An azimuthal single-spin asymmetry (SSA) arises from the convolution between TMD transversity distributions and the Collins fragmentation functions, which also contribute to the azimuthal correlations of nearly back-to-back hadrons in $e^+e^-$ annihilations. Joint analyses, including TMD evolution, find positive up-valence transversity, while the down flavor, sea quarks, and endpoint regions remain less constained~\cite{Kang:2015msa,Ye:2016prn,DAlesio:2020vtw,Zeng:2023nnb}.

Dihadron production in SIDIS provides access to transversity distributions within the collinear factorization. Here, transversity distributions couple to the chiral-odd interference dihadron fragmentation function (DiFF) $H_1^{\sphericalangle\,q}(z,M_h)$, which describes the fragmentation of a transversely polarized quark into a hadron pair, such as $\pi^+\pi^-$~\cite{Jaffe:1997hf,Radici:2001na,Bacchetta:2002ux,Metz:2016swz}. Together with the unpolarized DiFF $D_1^q(z,M_h)$, this function enters complementary observables in $e^+e^-$ annihilation, SIDIS, and polarized $pp$ collisions. Inclusive dihadron cross sections and azimuthal correlations between two hadron pairs in opposite hemispheres in $e^+e^-$ annihilation, as have been measured by Belle~\cite{Vossen:2011fk,Seidl:2017qhp}, provide constraint on the unpolarized and interference DiFFs. Transverse SSA in dihadron SIDIS, as have been measured by HERMES~\cite{Airapetian:2008sk} and COMPASS~\cite{Adolph:2012nw,Adolph:2014fjw,COMPASS:2023cgk,COMPASS:2025dwu}, provide flavor-sensitive combinations of transversity distributions and the interference DiFFs. Transverse SSA in inclusive dihadron production in $p^\uparrow p$ collisions, as have been measured by STAR~\cite{Adamczyk:2015hri,Adamczyk:2017ynk}, provide complementary constraints through convolutions involving transversity and the interference DiFF. Near-side energy-energy correlators provide a related approach within the dihadron fragmentation framework, with applications to SIDIS, $e^+e^-$ annihilation, and polarized $pp$ collisions~\cite{Kang:2026pro,Kang:2026qbm}.

Pioneering analyses extracted valence transversity from dihadron SIDIS data~\cite{Bacchetta:2011ip,Bacchetta:2012ty,Radici:2015mwa}, using interference DiFFs constrained by $e^+e^-$ measurements~\cite{Courtoy:2012ry}. Subsequent studies tested the universality of transversity against $pp$ data~\cite{Radici:2016lam} and incorporated into extractions~\cite{Radici:2018iag}. Nevertheless, the experimental constraints remain uneven in $x$ and flavor. The dihadron SIDIS data available for these analyses reach only $x \lesssim 0.3$, whereas the $pp$ data sample a broader range of partonic momentum fractions through convolutions involving various flavor channels. Consequently, the large-$x$ flavor decomposition and the full moments remain sensitive to the assumed functional form and extrapolation.

Lattice QCD provides complementary information through tensor charges, obtained from forward nucleon matrix elements of local quark tensor currents. In the parton picture, the tensor charges are the first Mellin moments of the valence transversity distributions,
\begin{equation}
  g_T^q(\mu)=\int_0^1 \dd{x}
  [h_1^q(x,\mu)-h_1^{\bar q}(x,\mu)].
  \label{eq:tensor-charge}
\end{equation}
They characterize the nucleon response to a tensor current, and also provide constraints on nonstandard tensor couplings in neutron $\beta$ decay~\cite{Bhattacharya:2011qm,Courtoy:2015haa} and on quark electric dipole moments~\cite{Liu:2017olr}. Lattice determinations of tensor charges have been used as constraints in phenomenological transversity extractions~\cite{Lin:2017stx,Gamberg:2022kdb}, including simultaneous determinations of transversity and DiFFs~\cite{Cocuzza:2023oam,Cocuzza:2023vqs}. By themselves, however, these integral constraints cannot distinguish distributions with the same moments but different $x$ dependence.

Lattice matrix elements of spatially separated quark fields provide access to the $x$ dependence of light-cone PDFs through large-momentum effective theory and related Euclidean approaches~\cite{Ji:2013dva,Ji:2014gla,Ji:2020ect,Radyushkin:2017cyf,Ma:2014jla,Ma:2017pxb,Cichy:2018mum,Constantinou:2020hdm}. Previous analyses of nucleon and pion structure have combined experimental data with lattice inputs, such as coordinate-space matrix elements, pseudo-Ioffe-time distributions, and generalized form factors~\cite{Bringewatt:2020ixn,JeffersonLabAngularMomentumJAM:2022aix,Karpie:2023nyg,Hunt-Smith:2024khs,Good:2025nny,Guo:2025muf,Huang:2026eai}. For transversity, lattice studies have examined the isovector flavor combination. Ref.~\cite{Yao:2022zmd} performed calculations at several lattice spacings and pion masses and extrapolated the results to the continuum limit and physical point. Another calculation used physical quark mass at a single lattice spacing~\cite{Gao:2023ktu}. At intermediate positive $x$, where the reconstruction is better controlled, the lattice data constrain $h_1^u-h_1^d$ over a wide range, both overlapping the upper end of the existing dihadron SIDIS measurements and extending to larger $x$. The negative-$x$ data provide constraints on the corresponding antiquark difference $h_1^{\bar u}-h_1^{\bar d}$.

In this Letter, we present the first determination of transversity distributions incorporating the $x$-dependent lattice data directly into a global analysis of dihadron production data. We extract transversity distributions and DiFFs simultaneously and compare matched neural-network (NN) fits with and without the lattice input to assess its impact on the flavor decomposition. Both fits describe the dihadron measurements comparably well. Including the lattice distribution shifts down-valence transversity toward more negative values and enhances up-valence transversity at large $x$. Integrating the valence distributions yields a larger isovector tensor charge, with most of the increase coming from the down-valence contribution shifted by the lattice data.

The Artru--Collins asymmetry in $e^+e^-$ annihilation, measured by Belle~\cite{Vossen:2011fk}, probes products of quark and antiquark interference DiFFs $H_1^\sphericalangle$ for hadron pairs produced in opposite hemispheres~\cite{Boer:2003cm}. In dihadron SIDIS on a transversely polarized target $N$, the interference DiFF $H_1^\sphericalangle$ couples to the quark transversity distribution $h_1$~\cite{Bacchetta:2002ux,Bacchetta:2004jz}. At leading twist and leading order in the hard scattering, the two asymmetries are given by
\begin{align}
 A_{e^+e^-}&=C_\theta
 \frac{\sum_q e_q^2\,
 H_1^{\sphericalangle\,q}(z_1,M_{h1})H_1^{\sphericalangle\,\bar q}(z_2,M_{h2})}
 {\sum_q e_q^2\,
 D_1^q(z_1,M_{h1})D_1^{\bar q}(z_2,M_{h2})},
 \label{eq:ee-asym}\\
 A_{UT}^{N}&=c_{\rm exp}(y)
 \frac{\sum_{q,\bar q} e_q^2\,h_1^{q/N}(x)H_1^{\sphericalangle\,q}(z,M_h)}
 {\sum_{q,\bar q} e_q^2\,f_1^{q/N}(x)D_1^q(z,M_h)}.
 \label{eq:sidis-asym}
\end{align}
For the Belle measurements, $C_\theta=\langle\sin^2\theta\rangle/ \langle1+\cos^2\theta\rangle$ is determined from the measured angular averages. In SIDIS, $y$ is the lepton inelasticity, and $c_{\rm exp}(y)$ accounts for the experimental asymmetry convention. It equals $1$ for COMPASS measurements~\cite{COMPASS:2023cgk} and $-(1-y)/(1-y+y^2/2)$ for HERMES measurements~\cite{Airapetian:2008sk}. In our analysis, we set the factorization scale $\mu=\sqrt{s}$ for $e^+e^-$ annihilation and $\mu=Q$ for SIDIS.

In $p^\uparrow p\to(\pi^+\pi^-)X$, transversity couples to the interference DiFF through transverse spin transfer in the hard scattering. STAR measures the amplitude $A_{UT}^{pp}$ of the $\sin\phi_{RS}$ modulation, with $\phi_{RS}=\phi_R-\phi_S$~\cite{Adamczyk:2015hri,Adamczyk:2017ynk}. This amplitude at leading order is given by a ratio of partonic convolutions~\cite{Bacchetta:2004it,Cocuzza:2023vqs}
\begin{equation}
 A_{UT}^{pp}=
 \frac{\begin{aligned}
  &\sum_{i,abcd}\int_{x_a^{\min}}^1 \dd{x_a}
   \int_{x_b^{\min}(x_a)}^1\frac{\dd{x_b}}{z}\\[-0.2em]
  &\quad\times h_1^a(x_a)f_1^b(x_b)\,
   \Delta\hat\sigma_i\,H_1^{\sphericalangle\,c}(z,M_h)
 \end{aligned}}
 {\begin{aligned}
  &\sum_{i,abcd}\int_{x_a^{\min}}^1 \dd{x_a}
   \int_{x_b^{\min}(x_a)}^1\frac{\dd{x_b}}{z}\\[-0.2em]
  &\quad\times f_1^a(x_a)f_1^b(x_b)\,
   \hat\sigma_i\,D_1^c(z,M_h)
 \end{aligned}},
 \label{eq:pp-asym}
\end{equation}
where $x_a$ and $x_b$ are the parton momentum fractions in the polarized and unpolarized protons, respectively. For a pion pair with transverse momentum $P_{hT}$ and pseudorapidity $\eta$, the fragmentation fraction is $z=\xi(e^\eta/x_a+e^{-\eta}/x_b)$, where $\xi=P_{hT}/\sqrt{s}$. The lower integration limits are $x_a^{\min}=\xi e^\eta/(1-\xi e^{-\eta})$ and $x_b^{\min}(x_a)=x_a\xi e^{-\eta}/(x_a-\xi e^\eta)$. The unpolarized and spin-transfer hard parts, $\hat\sigma_i=\dd{\hat{\sigma}_{ab\to cd}^{(i)}}/\dd{\hat{t}}$ and $\Delta\hat{\sigma}_i= \dd{\Delta\hat{\sigma}_{a^\uparrow b\to c^\uparrow d}^{(i)}}/\dd{\hat{t}}$, depend on the partonic invariants $\hat s=x_ax_bs$, $\hat t=x_at/z$, and $\hat u=x_bu/z$, with $t\simeq-\sqrt{s}P_{hT} e^{-\eta}$ and $u\simeq-\sqrt{s}P_{hT} e^\eta$. At fixed $P_{hT}$, the common factor $2P_{hT}$ cancels in the ratio. In the analysis, we set the factorization scale to $\mu=P_{hT}$ and evaluate the full partonic convolution for STAR measurements.

\begin{table*}
  \caption{Fit quality of the NN ensemble-mean predictions, reported as $\chi^2/N_{\rm dat}$ evaluated against the central values of the data. For LQCD, $N_{\rm dat}$ counts sampled $x$ points, and the ratio uses pointwise total uncertainties.
  \label{tab:data}}
  \begin{ruledtabular}
    \small
    \begin{tabular}{@{}llrrr@{}}
      Data set
      & Process
      & $N_{\rm dat}$
      & World data
      & World data + LQCD
      \\
      \hline
      Belle cross section~\cite{Seidl:2017qhp}
      & $e^+e^-\!\to(\pi\pi)X$ & 651 & 0.153 & 0.142 \\
      Belle asymmetry~\cite{Vossen:2011fk}
      & $e^+e^-\!\to(\pi\pi)(\pi\pi)X$ & 145 & 0.494 & 0.462 \\
      HERMES proton~\cite{Airapetian:2008sk}
      & $\ell p^\uparrow\!\to\ell'(\pi\pi)X$ & 12 & 0.886 & 0.785 \\
      COMPASS proton~\cite{COMPASS:2023cgk}
      & $\ell p^\uparrow\!\to\ell'(\pi\pi)X$ & 24 & 0.762 & 0.920 \\
      COMPASS deuteron~\cite{COMPASS:2023cgk}
      & $\ell d^\uparrow\!\to\ell'(\pi\pi)X$ & 24 & 0.804 & 0.834 \\
      STAR $200\,\mathrm{GeV}$~\cite{Adamczyk:2015hri}
      & $p^\uparrow p\!\to(\pi\pi)X$ & 24 & 1.172 & 1.239 \\
      STAR $500\,\mathrm{GeV}$~\cite{Adamczyk:2017ynk}
      & $p^\uparrow p\!\to(\pi\pi)X$ & 92 & 0.884 & 0.855 \\
      \hline
      All experiments
      & Combined world data & 972 & 0.338 & 0.329 \\
      Lattice QCD~\cite{Yao:2022zmd}
      & $h_1^{u-d}(x)$ from LQCD & 80 & --- & 0.423 \\
    \end{tabular}
  \end{ruledtabular}
\end{table*}

In this study, we perform a simultaneous analysis of the unpolarized dihadron production cross section~\cite{Seidl:2017qhp} and Artru--Collins asymmetry~\cite{Vossen:2011fk} in $e^+e^-$ annihilation, and the transverse SSAs in dihadron SIDIS~\cite{Airapetian:2008sk,COMPASS:2023cgk} and $pp$ collisions~\cite{Adamczyk:2015hri,Adamczyk:2017ynk}. We restrict DiFF kinematics in these processes to $0.25<z<0.95$ and $2m_\pi<M_h<1.20\,\mathrm{GeV}$. For the unpolarized dihadron production cross section, we also impose the collinearity cut $2M_h/(z\sqrt{s})<0.7$. The selected experimental data sample contains 972 points, as summarized in Table~\ref{tab:data}. With charge conjugation and isospin symmetry, the SIDIS data are sensitive to the valence combination $4h_1^{u_v}-h_1^{d_v}$ for the proton target and $h_1^{u_v}+h_1^{d_v}$ for the deuteron target. These data provide direct information on the flavor separation, but the highest bin-averaged $x$ value only reaches $0.287$.

To complement the flavor constraints by SIDIS data in limited region, we include the lattice determination of the isovector transversity distribution in the $\overline{\mathrm{MS}}$ scheme at $\mu=2\,\mathrm{GeV}$ from Ref.~\cite{Yao:2022zmd}. We sample the distribution at 40 points in each of the intervals $0.2\leq x\leq0.8$ and $-0.8\leq x\leq-0.2$. The positive-$x$ branch represents the quark distribution $h_1^{u-d}(x)=h_1^u(x)-h_1^d(x)$ and the negative-$x$ branch represents the antiquark distribution $h_1^{u-d}(-x)=-[h_1^{\bar u}(x)-h_1^{\bar d}(x)]$. Adding the positive and negative $x$ regions gives the isovector valence distribution $h_1^{u_v}(x)-h_1^{d_v}(x)$ and the normalization is restored from its integral by the isovector tensor charge $g_T^{u-d}=1.018(68)$. In the combined analysis, we retain the statistical correlations and coherent systematic uncertainties across $x$ through the Monte Carlo replicas, and the fit results are stable when varying the number of points sampled from the distribution, as tested numerically.

The extraction requires a joint representation of transversity distributions with $x$ and flavor dependence and the DiFFs with $(z,M_h)$ dependence at an input scale. Neural networks provide flexible multivariate representations~\cite{Hornik:1989} and have been used in PDF and DiFF analyses to reduce reliance on restrictive functional forms~\cite{Ball:2010de,Mahaut:2025diff}. Here we parametrize the transversity distributions and the DiFFs with fully connected NNs at $Q_0=1\,\mathrm{GeV}$, using hyperbolic-tangent hidden activations in the networks. For $\pi^+\pi^-$ production, we define the relative momentum as $R=(P_{\pi^+}-P_{\pi^-})/2$. Interchanging the pions reverses $R$ and the sign of the interference DiFF. For this ordering, charge conjugation and isospin symmetry give $D_1^u=D_1^d=D_1^{\bar u}=D_1^{\bar d}$ and $H_1^{\sphericalangle\,u}=-H_1^{\sphericalangle\,d}=-H_1^{\sphericalangle\,\bar u} =H_1^{\sphericalangle\,\bar d}$. Neglecting strange and heavy-quark components at the initial scale, we have two independent light-quark DiFFs, $D_1^u$ and $H_1^{\sphericalangle\,u}$. Our integrated $H_1^{\sphericalangle\,q}$ includes the pair-rest-frame factor $|\bm R|/M_h$ together with its angular weight~\cite{Pitonyak:2023gjx,Cocuzza:2023vqs}. All functions are evolved to the measurement scales via LO DGLAP evolution implemented in \mbox{HOPPET}~\cite{Salam:2008qg}.

The DiFF network takes $(z,M_h)$ as inputs and parametrizes the two independent light-quark functions $D_1^u$ and $H_1^{\sphericalangle\,u}$. We take three architectures with two hidden layers of $(32,32)$, $(48,48)$, and $(64,64)$. The transversity network takes $(x,\ln x)$ as inputs and parametrizes $h_1^{u_v}$, $h_1^{d_v}$, $h_1^{\bar u}$, and $h_1^{\bar d}$. We consider three architectures with two hidden layers of $(16,16)$, $(24,24)$, and $(32,32)$ nodes. The output layer is multiplied by $x^\alpha(1-x)^\beta$ and for antiquarks by additional factor of the corresponding unpolarized PDFs. While the NN outputs can compensate for these factors, the exponents $\alpha$ and $\beta$ remain model choices and may influence the small- and large-$x$ extrapolations beyond data coverage. We take nine endpoint choices, $(\alpha,\beta)\in\{0.8,1.2,1.6\}\times\{4,5,6\}$, in the analysis. Together with the architectures of the NNs, we have in total 81 parametrization models.

We assess the impact of the lattice input by comparing fits with and without the lattice distribution. Both fits use the same 200 Monte Carlo replicas of the experimental data, generated according to their uncertainties and correlations. For each replica, we randomly divide the data into training and validation sets. The training set guides the network updates, and the score used to select the checkpoint includes both sets. We combine the results from 81 parametrization models within each replica using exponential fit-score weights proportional to $\exp(-\Delta\chi^2/2)$, inspired by information-criterion averaging~\cite{Akaike:1974}. The weighted spread measures the sensitivity to the architectures and endpoint powers. The pointwise uncertainty of extracted physical quantities combines replica fluctuations with the weighted spread among parametrization models, accounting for data uncertainties and the model variations included in the ensemble.

Table~\ref{tab:data} reports $\chi^2/N_{\rm dat}$ for the NN ensemble-mean predictions evaluated against the central values of the data. The overall description of the experimental measurements remains comparable after including the lattice distribution. The lattice entry is evaluated using pointwise total uncertainties, whereas the fit retains correlations across $x$. This entry therefore only serves as a diagnostic of the agreement with input data.

\begin{figure}
  \centering  \resultgraphic{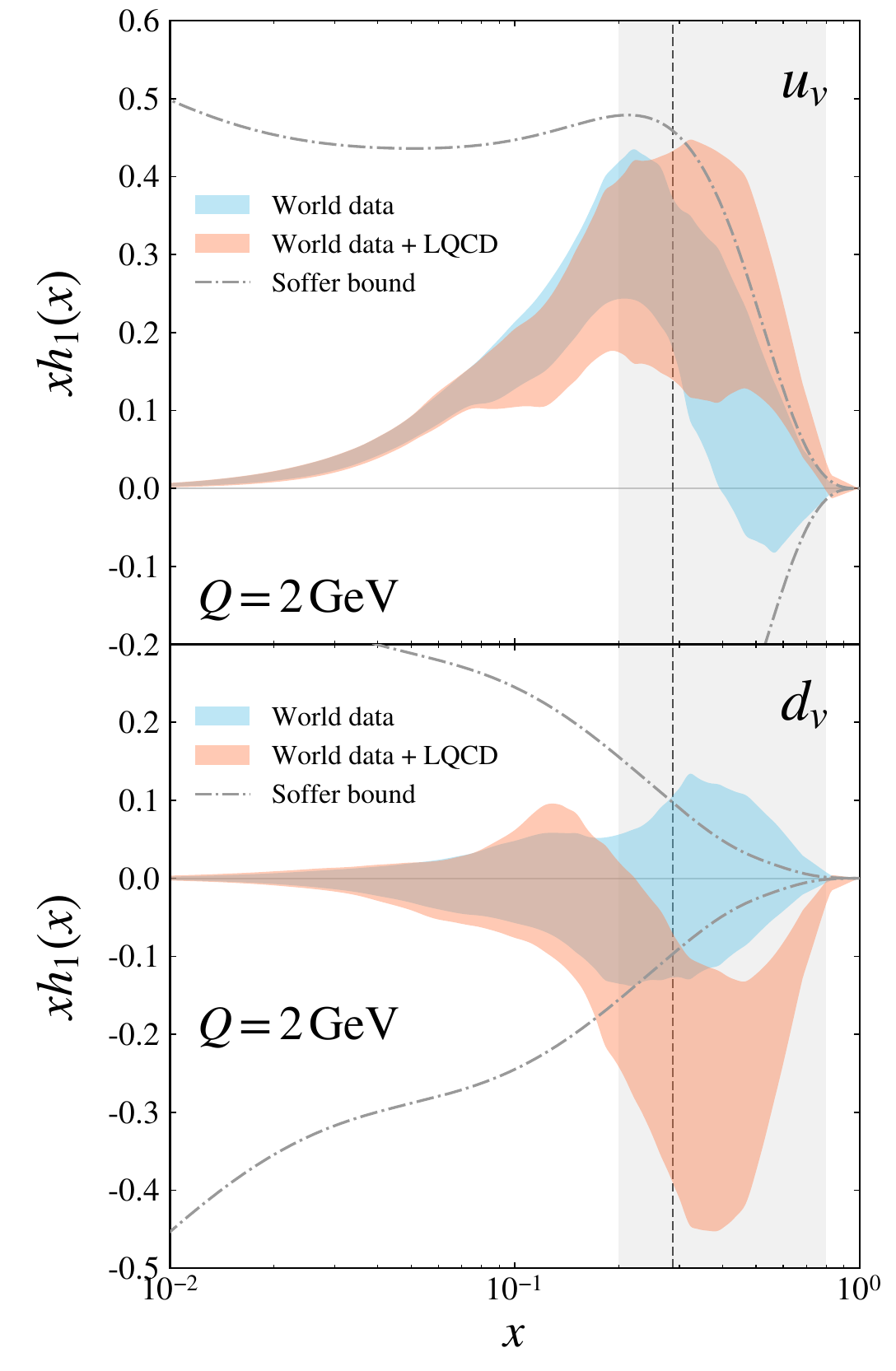}
  \caption{Extracted transversity distributions of $u_v$ (upper panel) and $d_v$ (lower panel) at $Q=2\,\mathrm{GeV}$. The uncertainty bands are shown at 68\% confidence level. The blue and orange bands are respectively the fits to experimental data alone and to experimental data combined with lattice QCD input. The vertical dashed line marks the highest bin-averaged $x$ value of the SIDIS data; gray shading marks the interval used for the lattice input. The gray dot-dashed curves represent the Soffer bound references, which are not imposed in the fit.
  \label{fig:transversity}}
\end{figure}

\begin{figure}
  \centering
  \resultgraphicat{0.98\linewidth}{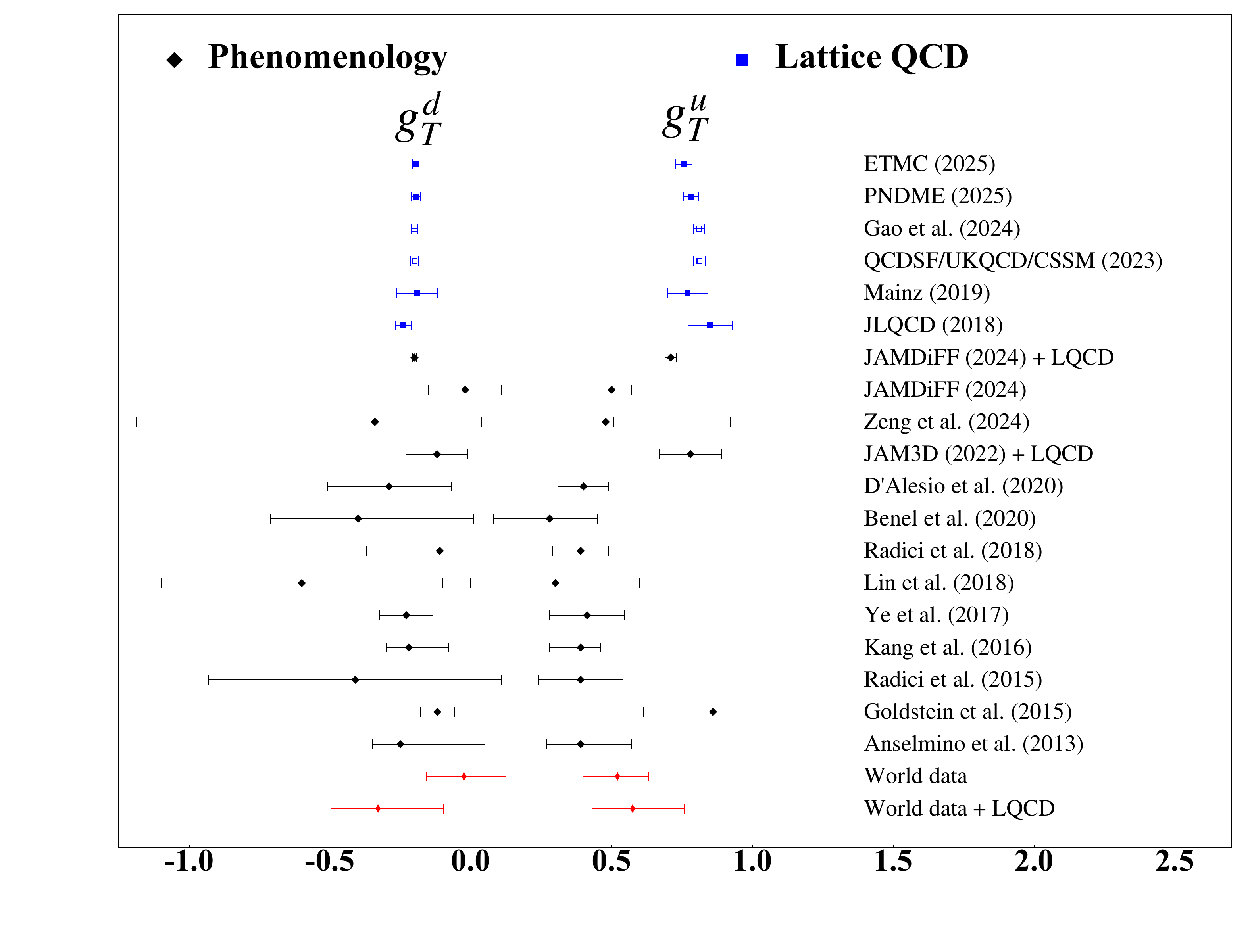}
  \par\vspace{0.25em}
  \resultgraphicat{0.98\linewidth}{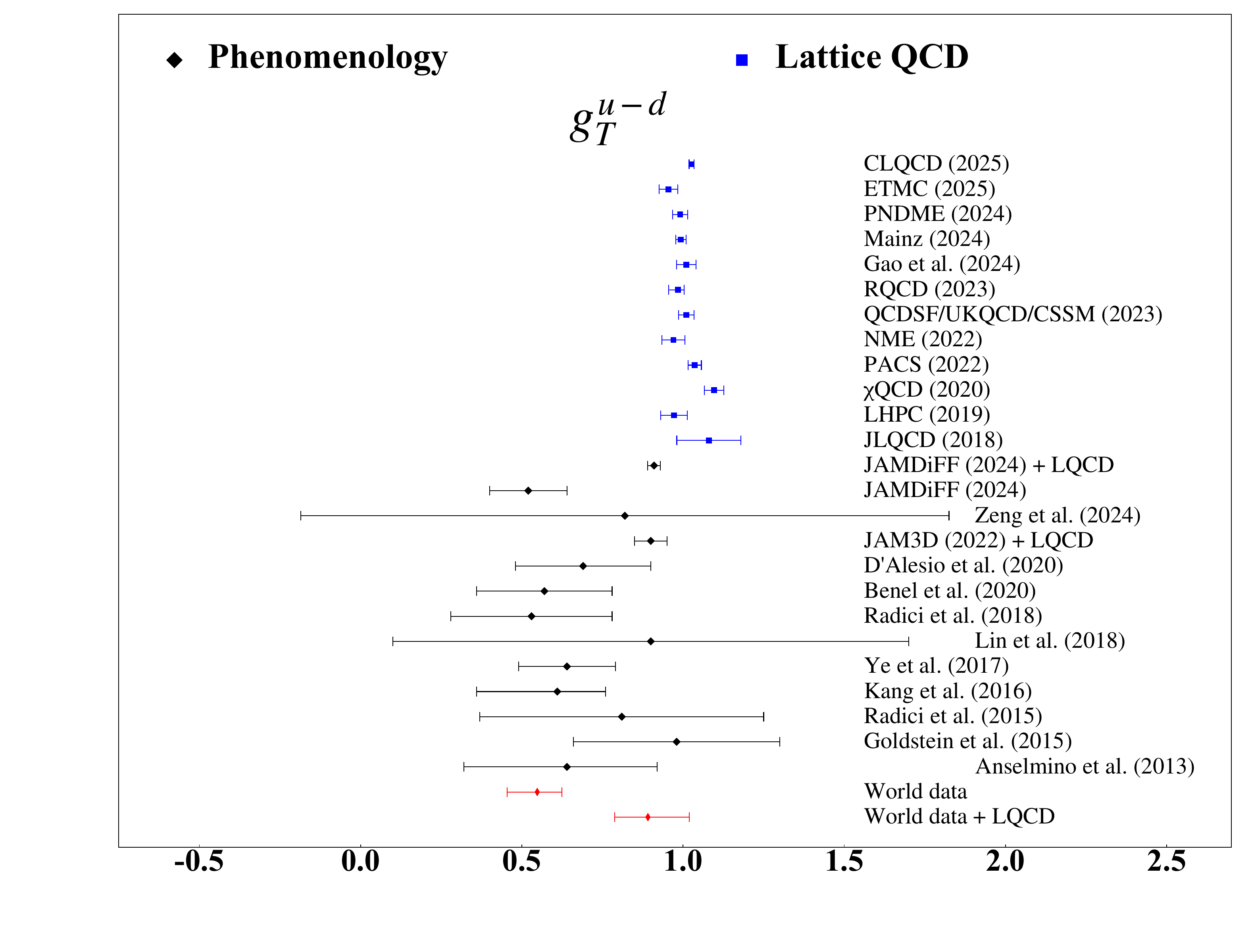}
  \caption{\label{fig:tensor}
  Flavor-diagonal (upper panel) and isovector (lower panel) tensor charges. Red diamonds denote results from our fits to experimental data alone (world-data) and to experimental data combined with lattice input (world-data + LQCD) at $Q=2\,{\rm GeV}$. Blue squares denote lattice determinations from Refs.~\cite{Yamanaka:2018uud,Hasan:2019noy,Djukanovic:2019gvi,Horkel:2020hpi,Tsuji:2022ric,Park:2021ypf,QCDSFUKQCDCSSM:2023qlx,Bali:2023sdi,Gao:2023ktu,Park:2025rxi,Djukanovic:2024krw,Jang:2023zts,Alexandrou:2024ozj,Wang:2025nsd}. Open squares denote connected-only flavor charges. Phenomenological comparisons include extractions using the dihadron mechanism~\cite{Radici:2015mwa,Radici:2018iag,Benel:2019mcq,Cocuzza:2023oam,Cocuzza:2023vqs}, the TMD Collins  effect~\cite{Anselmino:2013vqa,Kang:2015msa,Ye:2016prn,Lin:2017stx,DAlesio:2020vtw,Zeng:2023nnb}, the exclusive $\pi^0$ and $\eta$ electroproduction~\cite{Goldstein:2014aja,Courtoy:2015haa}, and analysis with lattice tensor charge constraints~\cite{Gamberg:2022kdb}.}
\end{figure}

As shown in Fig.~\ref{fig:transversity}, we compare the extracted valence transversity distributions with and without lattice input at $Q=2\,\mathrm{GeV}$. Experimental data favor positive $u_v$ transversity, while the $d_v$ transversity uncertainty band spans both positive and negative values over the full $x$ range. The lattice input provides complementary information on the isovector combination $h_1^{u_v}-h_1^{d_v}$ in a wide $x$ range. Together with the experimental constraints, this information shifts the central $d_v$ transversity distribution toward more negative values and enhances $u_v$ at large $x$, where direct SIDIS constraints are limited. These shifts in the distributions are accompanied by larger absolute uncertainties, reflecting the increased magnitudes of central values, but the relative uncertainties decrease at large $x$.

The gray dot-dashed curves in Fig.~\ref{fig:transversity} represent valence references derived from the quark and antiquark Soffer inequalities~\cite{Soffer:1994ww}, evaluated using central values of unpolarized and helicity PDFs~\cite{Yan:2022pzl,Cruz-Martinez:2025ahf}. These constraints are not imposed in the fit. The combined fit including lattice input yields a central $d_v$ transversity distribution that exceeds the corresponding reference in magnitude at large $x$, though with sizable uncertainties. Determining whether the experimental and lattice inputs can be jointly described while satisfying the Soffer inequalities motivates a constrained analysis that propagates the uncertainties of unpolarized and helicity PDFs. Further experimental measurements at high-$x$ at JLab~\cite{Sawaya:2026dqz} and the electron-ion colliders~\cite{AbdulKhalek:2021gbh,Anderle:2021wcy} are desired and would help clarify the flavor structure of transversity distributions.

Integrating the fitted valence distributions gives the proton tensor charges, as shown in Fig.~\ref{fig:tensor}. The fit to experimental data alone gives $g_T^u=0.521^{+0.112}_{-0.122}$, $g_T^d=-0.024^{+0.149}_{-0.133}$, and $g_T^{u-d}=0.548^{+0.076}_{-0.094}$. The fit including the lattice input gives $g_T^u=0.574^{+0.185}_{-0.144}$, $g_T^d=-0.329^{+0.231}_{-0.167}$, and $g_T^{u-d}=0.891^{+0.129}_{-0.103}$. The quoted values are given at $Q=2\,{\rm GeV}$ with uncertainties at 68\% confidence level. The increase in $g_T^{u-d}$ is driven mainly by the more negative down-valence contribution, bringing the central values closer to the lattice determinations shown in Fig.~\ref{fig:tensor}. The closer agreement should be interpreted in light of the tensor-charge normalization already incorporated via the lattice input distribution. The additional information supplied by the $x$ dependence constrains the flavor shapes, while the full moments still depend on extrapolation beyond the fitted $x$ interval, despite the broader $x$ coverage provided by the lattice input.

In summary, we present the first simultaneous determination of collinear transversity distributions and DiFFs combining world dihadron production data and $x$-dependent lattice calculations. The lattice input supplies information on quark and antiquark flavor differences beyond tensor charges alone, complementing the experimental flavor sensitivity, particularly at large $x$ where SIDIS data constraints are limited. The fit including lattice input favors a more negative down-valence transversity distribution and enhanced up-valence transversity distribution at large $x$, while preserving a comparable description of the experimental data. The more negative down-valence contribution drives most of the increase in the isovector tensor charge. These results demonstrate the complementary role of experimental measurements and lattice information on the $x$ dependence in resolving transversity distributions, providing a basis for further joint studies of nucleon spin structure.

\acknowledgments{
This work was supported by the National Natural Science Foundation of China (NSFC) under Grants No.~12321005, No.~12322503, and No.~12675120, the National Key R\&D Program of China No.~2024YFA1611004, and the CAS Project for Young Scientists in Basic Research No.~YSBR-117. The work of H. D. was also supported by the Postgraduate Research \& Practice Innovation Program of Jiangsu Province under Grant No.~KYCX25\_1936.
}

\bibliography{reference}

\end{document}